\documentclass[onecolumn,final,numbook]{svjour22}
\usepackage{graphicx}
\usepackage{mathptmx}
\usepackage{amssymb}
\usepackage{amsfonts}
\usepackage{amsmath}

\renewcommand{\v}{\vec{v}} 
\newcommand{\X}{\vec{X}} 
\newcommand{\Y}{\vec{Y}} 
\newcommand{\x}{\vec{x}} 
\newcommand{\y}{\vec{y}} 
\newcommand{\R}{\vec{R}} 
\newcommand{\F}{\mathcal{F}} 
\newcommand{\G}{\mathsf{F}} 
\renewcommand{\P}{\mathcal{P}} 
\renewcommand{\d}{\mathrm{d}}
\renewcommand{\dim}{d}
\newcommand{\dd}{\mathrm{d}^\dim}
\newcommand{\mean}[1]{\left\langle #1\right\rangle} 
\newcommand{\dert}[1]{\frac{\d{#1}}{\d t}}
\journalname{J Stat Phys}

\usepackage{color}

\begin{document}

\title{Effective rates in dilute reaction-advection systems for the
  annihilation process $A+A\to \varnothing$}

\author{G. Krstulovic \and M. Cencini \and J. Bec}

\institute{Giorgio Krstulovic \and J{\'e}r{\'e}mie Bec \at Laboratoire
  Lagrange, UMR7293, Universit\'e de Nice Sophia-Antipolis, CNRS,
  Observatoire de la C\^ote d'Azur, BP 4229, 06304 Nice Cedex 4,
  France \and Massimo Cencini \at
  CNR, Istituto dei Sistemi Complessi, Via dei Taurini 19, Roma,
  Italy}

\date{\today}

\maketitle

\begin{abstract}
  A dilute system of reacting particles transported by fluid flows is
  considered. The particles react as $A+A\to \varnothing$ with a given
  rate when they are within a finite radius of interaction. The system
  is described in terms of the joint $n$-point number spatial density
  that it is shown to obey a hierarchy of transport equations. An
  analytic solution is obtained in the dilute or, which is equivalent,
  the long-time limit by using a Lagrangian approach where statistical
  averages are performed along non-reacting trajectories. In this
  limit, it is shown that the moments of the number of particles have
  an exponential decay rather than the algebraic prediction of
  standard mean-field approaches. The effective reaction rate is then
  related to Lagrangian pair statistics by a large-deviation
  principle. A phenomenological model is introduced to study the
  qualitative behavior of the effective rate as a function of the
  interaction length, the degree of chaoticity of the dynamics and the
  compressibility of the carrier flow.  Exact computations,
    obtained via a Feynman--Kac approach, in a smooth, compressible,
  random delta-correlated-in-time Gaussian velocity field support the
  proposed heuristic approach.
\end{abstract}

\section{Introduction}
\label{intro}

Many natural and industrial processes involve the reaction or
collision of diffusing species transported by an outer flow.  Such
systems are typically modeled in terms of the
\emph{reaction-diffusion-advection} equation for the density field
$\rho$. In the simple case of the pair-annihilation reaction $A+A
\rightarrow\varnothing$, where two molecules react together to become
inert, this kinetic equation takes the form
\begin{equation}
  \partial_t \rho + \nabla\cdot(\rho\v) = -\Gamma\,\rho^2 + \kappa
  \nabla^2\rho, 
  \label{eq:adv-reactGen}
\end{equation}
where $\kappa$ is the diffusion constant, $\v$ the velocity of the
outer flow, and $\Gamma$ the reaction rate. An important aspect
withheld in such an approach is the assumed relation between the
microscopic stochastic rate, that is the probability that two given
individual molecules react, and the mesoscopic reaction propensity,
that is the number of reactions per unit time and volume written here
as $\Gamma\,\rho^2$. To derive \eqref{eq:adv-reactGen} one must assume
two properties to be satisfied at the coarse-graining scale from which
the hydrodynamic limit is taken \cite{vanKampen1992}. First, volumes
at the coarse-graining scale have to contain sufficiently many
particles, in order to safely disregard finite-number
fluctuations. Second, each particle within this volume must have an
equal probability to react with all the others --- \emph{well-mixing}
hypothesis.  For this second condition to be satisfied, the
coarse-graining scale has to be sufficiently small, i.e.\ smaller than
the interaction distance between particle pairs.

It is clear that in situations where the reactants are very dilute,
the two conditions underlying \eqref{eq:adv-reactGen} might not be
simultaneously satisfied. In particular, fluctuations due to a finite
number of reactants might be so important to invalidate the mean-field
assumptions leading to \eqref{eq:adv-reactGen}.  Much work has been
devoted to model and study such fluctuations in situations where
transport is negligible and diffusion dominates. It was shown that
finite-number effects can be taken into account by adding an imaginary
noise in the reaction-diffusion equation. The time evolution of
statistical quantities is then obtained by averaging with respect to
this noise (see~\cite{mattis1998uses,Tauber2005} for reviews). Two
equivalent approaches were independently developed using either the
Poisson representation~\cite{gardiner1977poisson} or a
field-theoretical
description~\cite{doi1976stochastic,peliti1985path}. In both cases, it
is assumed that a reaction takes place with a given rate when two
particles are on the same node of a lattice. Diffusion is then modeled
by a random walk. After that, the hydrodynamic limit is obtained by
performing the limit of vanishing lattice spacing. Such approaches,
which were developed for diffusion-limited reactions, cannot be
straightforwardly extended to cases where transport cannot be
neglected.

In many natural situations, the dynamics of very dilute reacting
species is dominated by (possibly compressible) advection and their
diffusion is almost negligible. This is the case, e.g., of dust grains
growing by accretion to form planets in the early solar
system~\cite{johansen2007rapid} or of phytoplankton confined in a
two-dimensional ocean layers~\cite{Perlekar2010}. Also, motile
microorganisms in aquatic environments can detach from the fluid
trajectories thanks to swimming and behave as tracers in a
weakly-compressible flow \cite{Neufeld2007}. Because of the large
sizes of the transported species, molecular diffusion is negligible in
all aforementioned examples. Another instance concerns the initiation
of rain in warm clouds. Droplets, which are typically occupying a
volume fraction of the order of $10^{-6}$, grow by coalescence to form
raindrops. Because of their finite size and mass, such droplets have
inertia and their dynamics decouples from the flow. When they are
sufficiently small, droplets behave as tracers in an effective
compressible flow that depends on the local turbulence of the carrier
airflow~\cite{Maxey2006}. This leads to strong and violent
fluctuations both in their local spatial
concentration~\cite{shaw2003particle} and in the rate at which they
collide~\cite{falkovich2002acceleration}. Classical approaches consist
in using the (mean-field) Smoluchowski coagulation equation, which is
a generalization of the kinetic equation~\eqref{eq:adv-reactGen} to
the case of a full population with a size
distribution~\cite{pruppacher1998microphysics}. Many questions remain
open on the statistical effects of turbulence on the timescales at
which rain is formed~\cite{devenish2012droplet}. In addition very
little is known on the finite-number fluctuations induced by
diluteness that might also play an important role in assessing the
average growth rate of droplets. The aim of this work is to develop a
general framework which can be used to address such issues. For that,
we neglect diffusion and focus on particles transported by generic
compressible flows. Also, we focus on the annihilation reaction
$A+A{\rightarrow}\varnothing$ because we expect that this simple model
will capture the essence of finite-number effects in
advection-reaction systems.

It is clear that, at long times, the particle number density
decreases, so that this long-term asymptotics is very likely affected
by finite-number effects. Kinetic approaches using
\eqref{eq:adv-reactGen} predict that, ultimately, the average number
of particles decays algebraically as
$t^{-1}$~\cite{chertkov2003boundary}.  This behavior, which occurs at
times much longer than the fluid velocity correlation time, can be
explained in terms of an effective eddy diffusion. The flow
compressibility might decrease the effective
diffusivity~\cite{vergassola1997scalar}, but the latter remains anyway
positive~\cite{goudon2004homogenization}. Hence, at long times, the
spatial fluctuations of the concentration are smoothed out and a
closed equation can be written for the spatial average of
$\rho$. Here, we find that finite-number effects enhance the
long-time decay of the average number of particles. For that we
consider discrete particles that are tracers of the compressible fluid
flow and that annihilate with a rate $\mu$ when they are separated by
less than an interaction distance $a$. We show that the average number
of particles does not decrease algebraically but rather exponentially
as $\exp(-\gamma\,t)$. This law pertains to the statistics of the
relative motion between two reactants and the exponential decay rate
$\gamma$ depends on both the microscopic rate $\mu$ and the flow
statistical properties in a non-trivial manner. The main result of
this work is the introduction of a novel Lagrangian approach in terms
of non-reacting particle trajectories. We exploit these ideas to
express $\gamma$ using a large-deviation principle for the time that
two tracers spend at a distance below the radius of interaction $a$.

The paper is organized as follows. Section \ref{sec:master} contains
basic definitions and set the general framework of this work. The
$n$-point number-density field is introduced and is shown to obey a
hierarchy of transport equations. When integrated over space, these
fields correspond to the factorial moments of the number of particles
that are present in the system. We then briefly discuss the
zero-dimensional case and its relationship with standard studies of
finite-number effects in well-mixed settings.
In section \ref{sec:lagrange}, the hierarchy for the $n$-point
number-density is solved by using a Lagrangian approach that consists
in following the flow characteristics. At long times the particle
number moments are shown to decay exponentially with a rate $\gamma$
that does not depend on their order.  An analytic expression for
$\gamma$ is given in terms of the Lagrangian statistics of
(non-reacting) tracer trajectories.
In section \ref{sec:2point} we focus on the relevant case of two
particles. The exponential decay rate is then related to the time
spent by the pair at a distance less than the interaction radius and
is expressed in terms of the large fluctuations of the
latter. Asymptotic arguments and a phenomenological picture valid for
generic flows are then developed to relate the decay rate $\gamma$ to
the \emph{microscopic} rate $\mu$.
Section \ref{sec:dilute_kraich} contains an application to very dilute
suspensions in smooth compressible Gaussian delta-correlated-in-time
flows (belonging to the so-called \emph{Kraichnan ensemble}; see,
e.g., \cite{FalkovichG.2001}) for which the effective reaction
  rate can be analytically derived.
Finally, section \ref{sec:conclusion} presents some concluding remarks
and open questions.

\section{Master equations in the presence of advection\label{sec:master}}

\subsection{Settings\label{ssec:settings}}

We consider a set of particles $\{A_i\}_{1\le i\le N(t)}$ whose
positions $\X_i$ obey the equations
\begin{equation}
  \dert{\X_i} = \v(\X_i,t),
  \label{eq:tracers}
\end{equation}
where $\v$ is a prescribed $d$-dimensional differentiable velocity
field with given isotropic, stationary, and homogeneous
statistics. 
The velocity field $\v$ is assumed to have a finite
correlation time and might be compressible with
\emph{compressibility} 
\begin{equation}
  \wp =
  \frac{\overline{\left(\nabla\cdot\v\right)^2}}
  {\ \overline{\mbox{tr}\left(\nabla\v^\mathsf{T}\,\nabla\v\right)}\ },
\end{equation}
where the over line designates the average with respect to the
velocity field realizations, and $\wp\in[0:1]$. The extremal values
$\wp=0$ and $\wp=1$ correspond to incompressible and purely potential
velocity fields, respectively. Whenever $\wp>0$, the dynamical system
defined by \eqref{eq:tracers} is dissipative. Also, to maintain
sufficiently generic settings, the dynamics is assumed ergodic and
chaotic. Typically, when the flow is defined on a compact set, the
trajectories generated by \eqref{eq:tracers} will concentrate on a
dynamically evolving strange attractor, while in the incompressible
limit they remain uniformly distributed (see, e.g.,
\cite{FalkovichG.2001} for more details on particle transport in
compressible and incompressible flows). For the sake of simplicity, in
this paper we will focus on the dynamics of tracers given by
\eqref{eq:tracers}. However, the results can be straightforwardly
extended to a general dynamics given by the Newton equation
$\ddot{\X}={\bf F}(\X,\dot{\X},t)$ as that ruling, for instance, the
evolution of inertial particles \cite{Maxey2006}.

The dynamics \eqref{eq:tracers} is supplemented by binary reactions
between the particles.  When $|\X_i - \X_j|<a$ the particles labeled
$i$ and $j$ might react and annihilate (or become inert). This happens
with a rate $\mu$. As a consequence, the number of particles $N(t)$ in
the domain will decrease from its initial value $N(0) = N_0$ in the
course of time. We suppose that the radius of interaction $a$ is
smaller than the scale at which $\v$ varies.  For example, in
turbulent flows, this corresponds to assuming that $a$ is smaller than
the Kolmogorov length scale $\ell_\mathrm{K}$.  Moreover, in writing
\eqref{eq:tracers} we have neglected particle diffusion. This is
justified whenever the interaction distance $a$ is larger than the
Batchelor length scale $\ell_\mathrm{B}= \ell_\mathrm{K}/\sqrt{Sc}$,
where $Sc = \nu/\kappa$ designates the Schmidt number --- that is the
ratio between the fluid kinematic viscosity $\nu$ and the particle
diffusivity $\kappa$. For applications, our settings thus reduce to
very large values of the Schmidt number. We notice that the Schmidt
number can be of the order of thousands or higher in organic mixtures,
biological fluids and generically for particulate matter. Estimates
based on Stokes--Einstein relation lead for a micro-meter sized
spherical particle $Sc\approx1000$ in air and $10^6$ in water.

Before discussing the master equations which describe the system, it
is worth underlining that our settings involve different sources of
stochasticity. First, the dynamics \eqref{eq:tracers} has to be
supplemented by initial conditions on particle positions that we
choose randomly. A second source of randomness comes from the
intrinsic stochasticity of the reaction process. Finally, the
  fluid flow is stochastic with prescribed statistics.  Consequently,
ensemble averages corresponding to these different sources of
randomness must be considered; the corresponding notations will be
introduced when needed. 

\subsection{Master equations for $n$-point number densities}
\label{ssec:master}

To fully describe all interactions between the particles in the
  considered system, all relative positions are simultaneously
  needed. It is then natural to define for
$1\le n\le N_0$ ($N_0=N(0)$ being the initial number of particles) the
joint $n$-point number density
\begin{equation}
  \F_n(\x_1,\ldots,\x_n,t)=\left\langle\sum_{i_1\neq\ldots\neq i_n}
    \delta\left(\X_{i_1}(t)-\x_1\right)\cdots\delta
    \left(\X_{i_n}(t)-\x_n\right)\right\rangle_\mu, \label{Eq:defPn} 
\end{equation}
where the brackets $\langle\cdot\rangle_\mu$ stands for the ensemble
average with respect to reactions only and, by convention, the sum in
\eqref{Eq:defPn} is zero when $N(t)<n$. The particle indices $i_k$
vary between $1$ and the number $N(t)$ of particles in each
realization of the reactions. The sum is not
ordered so that, by construction, the spatial averages of the $\F_n$'s
are the factorial moments of the total number of particles, namely
\begin{eqnarray}
  {\G}_1(t)&=&\int \F_1(\x_1,t)\,\dd x_1=\mean{N(t)}_\mu, \\
  {\G}_2(t)&=&\iint \F_2(\x_1,\x_2,t)\,\dd x_1\,\dd x_2 =
  \mean{N(t)\,(N(t)-1)}_\mu\\
  {\G}_n(t)&=&\idotsint \F_n(\x_1,\ldots,\x_n,t)\,\dd x_1\cdots
  \dd x_n=\mean{\frac{N(t)!}{(N(t)-n)!}}_\mu\,.\label{Eq:defFactMoments}
\end{eqnarray}
Hence, ${\G}_n(t)/n!$ is the average number of $n$-uplets among the
$N(t)$ particles still present at time $t$.

Taking into account the transport by the flow $\v$ and the reactions,
the joint $n$-point densities \eqref{Eq:defPn} satisfy the following
hierarchy of equations
\begin{eqnarray}
  \frac{\partial \F_n}{\partial t}+\sum_{i=1}^n
  \nabla_{\x_i}\cdot[\v(\x_i,t)\,\F_n] &=& -\mu\,
  \F_n\,\sum_{i<j}\theta(a-|\x_i-\x_j|) \nonumber \\ && \!\!\!\!\!\!\!\!\!\!\!
  -\mu
  \sum_{i=1}^n\int \limits \theta(a-|\x_i-\x_{n+1}|) \F_{n+1} \,\dd x_{n+1},
  \label{Eq:DefMasterEq}
\end{eqnarray}
where $\nabla_{\x_i}\cdot$ is the $d$-dimensional divergence with
respect to the spatial variable $\x_i$ and $\theta$ the Heaviside
function. The transport term comes from the identity
\begin{equation}
\dert{}\delta(\X_{i_k}(t)-\x_k)=-\nabla_{\x_k} \cdot\left[
  \delta(\X_{i_k}(t)-\x_k)\, \v(\x_k,t) \right].
\end{equation}
The two terms in the right-hand side are sinks accounting for the
reactions: the first includes all binary reactions between any
close-enough pair among the $n$ considered particles; the second
counts the reactions of any of these $n$ particles with another one at
a distance smaller than $a$.

The hierarchy \eqref{Eq:DefMasterEq} for the $n$-point joint number
density is exact and contains all information about the dynamics and
statistics of the system. Moreover, it is closed at the order $n=N_0$
and can be integrated by the method of characteristics (see
Sect.~\ref{sec:lagrange}). In the next subsection we relate this
description in terms of factorial moments to standard descriptions
used in the statistical physics of well-mixed chemical reactions.

\subsection{The zero-dimensional case}
 
When the hypothesis of \emph{well-mixedness} is satisfied, as when
  for instance the interaction radius is larger than the system size,
transport does not play any role in the temporal evolution of the
factorial moments $ {\F}_n$ \eqref{Eq:defPn}. In this zero-dimensional
limit, the hierarchy of transport equations \eqref{Eq:DefMasterEq}
becomes a simpler hierarchy of ordinary differential equations, i.e.
\begin{equation}
  \frac{\d\G_n}{\d t}=-\frac{\mu}{2}\,n\,(n-1)\, \G_n
  -\mu\,n\,\G_{n+1}\,,
  \label{Eq:DefMasterEq0D} 
\end{equation}
which can be also derived from the standard master equation describing
the reaction $A+A{\longrightarrow} \varnothing$. Denoting with
$\P_N(t)$ the probability of having $N$ particles at time $t$, we have
\cite{vanKampen1992}
\begin{equation}
  \dert{\P_N}=\frac{\mu}{2}\,(N+2)\,(N+1)\,\P_{N+2}
  -\frac{\mu}{2}\,N\,(N-1)\,\P_N.
  \label{Eq:MasterEqf}
\end{equation}
By definition, the factorial moments are expressed in terms of $\P_N(t)$ as
\begin{equation}
\G_n(t)=\sum_{N=0}^\infty \frac{N!}{(N-n)!}\, \P_N(t).
\end{equation}
Taking the time derivative of $\G_n(t)$ and using \eqref{Eq:MasterEqf}
yields to \eqref{Eq:DefMasterEq0D}. 

Equation~\eqref{Eq:MasterEqf} is generally the starting point for
describing the statistics of chemical reactions. In the limit of large
number of particles, the mean-field approximation $\langle N(t)
(N(t)-1)\rangle_\mu\approx \langle N(t)^2\rangle_\mu\approx\langle
N(t)\rangle_\mu^2$ is often used to write from\eqref{Eq:MasterEqf} a
closed equation for the mean number of particles. This mean-field
equation, obtained by replacing $\G_2$ by $\G_1^2$ in
\eqref{Eq:DefMasterEq0D} for $n=1$, predicts a $t^{-1}$ algebraic
decay of the number of particles. Clearly, the mean-field
approximation will breakdown at long times when the number of
particles is smaller.  Many works have been devoted to describe
fluctuations and deviations from mean field due to a finite number of
particles \cite{CardyImaginaryNoise,Deloubriere2002,Tauber2005}. In
particular, it is possible to show that solving equation
\eqref{Eq:DefMasterEq0D} is equivalent to computing the moments of the
solution of a stochastic differential equation with a pure imaginary
noise~\cite{CardyImaginaryNoise,Gardiner_1985}.  However, this
approach cannot be directly applied when the interaction radius is
finite and transport is present. Reactions and spatial location of
particles cannot be treated separately as it is usually done in
reaction-diffusion systems, where the \emph{well-mixed} hypothesis is
assumed at the scale of the lattice spacing. The hierarchy of
transport equations \eqref{Eq:DefMasterEq} naturally couples the
spatial distribution of particles and the reactions. The aim of the
next section is to provide a formal solution of \eqref{Eq:DefMasterEq}
in terms of Lagrangian trajectories (or tracers) of the carrier
flow. These trajectories are completely defined by carrier flow and do
not depend on the reactions.

\section{A Lagrangian approach to particle reaction kinetics}
\label{sec:lagrange}

\subsection{Finite-number closure and recursive solutions}
\label{ssec:notitle}
We now consider a finite interaction distance so that the presence of
transport by the flow must be explicitly accounted for.  Without loss
of generality, we can assume that at $t=0$ the total number of
particles is fixed and equal to $N_0$.  As $\F_n=0$ for all $n>N_0$,
the hierarchy \eqref{Eq:DefMasterEq} is closed at the order $N_0$ and
can be integrated.

Let us start with the joint $N_0$-point density
\begin{equation}
  \frac{\partial \F_{N_0}}{\partial t}+
    \sum_{i=1}^{N_0}\nabla_{\x_i} \cdot[\F_{N_0}\,\v(\x_i,t)] =-\mu
    \F_{N_0} \sum_{i<j\le N_0}\theta(a-|\x_i-\x_j|),
    \label{Eq:DefMasterEqN0}
\end{equation}
which can be formally solved in term of Lagrangian trajectories ---
that is using the method of characteristics. Denoting by $\Y(t;\y)$
the solution to the differential equation
\begin{equation}
\frac{\d }{\d t}\Y(t; \y)=\v({\Y} (t; \y),t) \,,\,\,\,\Y(0;\y)
=\y\label{Eq:defLagTraj}\,,
\end{equation}
it is straightforward to check
that the solution of \eqref{Eq:DefMasterEqN0} is given by
\begin{eqnarray}
 &&   \F_{N_0}(\x_1,\ldots,\x_{N_0},t)  = \nonumber\\ 
  && = \!\int
  \F_{N_0}^0\,\mathrm{e}^{ -\mu\sum_{i<j\le  N_0}
    \int_0^t\theta(a-|\Y(s;\y_i)-\Y(s;\y_j)|) \d s}
  \prod_{i=1}^{N_0}\,\delta(\Y(t;\y_i)-\x_i)\, \d 
  y_1\cdots \d y_{N_0} \nonumber \\
  &&\equiv \left\langle \F_{N_0}^0\,\mathrm{e}^{-\mu\sum_{i<j\le
        N_0} \int_0^t\theta(a-|\Y(s;\y_i)-\Y(s;\y_j)|) \,\d
      s}\right\rangle_{\!\!\!\!N_0}\!\!,
\end{eqnarray}
with
$\F_{N_0}^0(\y_1,\dots,\y_{N_0})=\F_{N_0}(\y_1,\dots,\y_{N_0},0)$. We
have introduced here the so-called Lagrangian average $\langle \cdot
\rangle_{N_0}$, which is an average over all $N_0$-uplets of
non-reacting trajectories such that their location at time $t$ is
$\x_1,\dots,\x_{N_0}$. Such an average is often used in compressible
transport \cite{gawedzki2000phase}. We can now insert the expression
for $\F_{N_0}$ in the equation for $\F_{N_0-1}$ obtaining
\begin{eqnarray}
  \frac{\partial \F_{N_0-1}}{\partial
    t} + \sum_{i=1}^{N_0-1}\!\! \nabla_{\x_i}\cdot[\F_{N_0-1} \v(\x_i,t)]
  =  -\mu \F_{N_0-1} \sum_{i<j}^{N_0-1}\theta(a-|\x_i-\x_j|)
  \nonumber\\
  \quad-\mu \left\langle \F_{N_0}^0\,
    \mathrm{e}^{-\Lambda_{N_0}(t)} \sum\limits_{i=1}^{N_0-1}
      \theta(a-|\Y(t;\y_i)-\Y(t;\y_{N_0})|) \right\rangle_{\!\!\!\!N_0}\!\!,
\label{Eq:DefMasterEqN0m1}
\end{eqnarray}
where
\begin{equation}
  \Lambda_n(t)=\mu\sum\limits_{i<j\le
    n}\int_0^t\theta(a-|\Y(s;\y_i)-\Y(s;\y_j)|)\,\d s.
\end{equation}
The solution of Eq.\eqref{Eq:DefMasterEqN0m1} reads
\begin{eqnarray}
 &&\F_{N_0-1}(\x_1,\ldots,\x_{N_0},t)=\mean{\F_{N_0-1}^0\,
   \mathrm{e}^{-\Lambda_{N_0-1}(t)}}_{\!\! \!\!N_0-1}
 \label{Eq:SolPNm1}\\ &&
 -\mu\!\mean{\!\!\F_{N_0}^0\,\mathrm{e}^{-\Lambda_{N_0-1}(t)}\!
   \sum\limits_{i=1}^{N_0-1}\!\!\int_0^t\!\! \theta(a-|\Y(s;\y_i)\!
   -\!\Y(s;\y_{N_0})|)\,\mathrm{e}^{\Lambda_{N_0-1}(s)-\Lambda_{N_0}(s)
   } \d s\!}_{\!\!\!\!N_0}\!\!\!\!\!\!. \nonumber
\end{eqnarray}

The general solution for $\F_{n}$ can be obtained reintroducing
\eqref{Eq:SolPNm1} in \eqref{Eq:DefMasterEq} solving for $\F_{N-2}$
and iterating until the order $n$.

In writing \eqref{Eq:defLagTraj} we have implicitly assumed the
existence and uniqueness of the Lagrangian trajectories and each
trajectory is labeled by its initial position. This assumption is
guaranteed as we have considered a smooth velocity field.  For a
  fully turbulent flow and if $a$ is much larger than the Kolmogorov
  length scale, the velocity field is no longer smooth but only
  H\"older continuous with an exponent smaller than one. In such a
case the solutions to \eqref{Eq:defLagTraj} are not unique and the
deterministic Lagrangian flow breaks down. However, a statistical
description of Lagrangian trajectories is still possible
\cite{gawedzki2000phase}.

We conclude this subsection by noticing that this new approach in
terms of Lagrangian trajectories allows us to directly apply known
results from passive (non-reacting) transport (see
\cite{FalkovichG.2001} for a review) to address the problem of
advection-reaction. These theoretical considerations will be mostly
used in Sect.~\ref{sec:2point}.

\subsection{Long times and very dilute closures \label{ssec:longtime}}

When considering very dilute systems and small interaction radii, the
hierarchy of transport equations can be closed at each order as the
second term of the right hand side of Eq.~\eqref{Eq:DefMasterEq} can
be neglected with respect to the first one. In particular, the closure
becomes exact in the long time limit when the mean number of
particles depends only on the statistics of the relative distance of
one pair of Lagrangian trajectories.

The long time behavior of the joint $n$-point density can be directly
obtained from equation \eqref{Eq:SolPNm1}. Noticing that
$\Lambda_{N_0-1}(s)-\Lambda_{N_0}(s)<0$, at the leading order, the
joint $n$-point density is given for $n\ge2$ by
\begin{equation}
  \F_{n}(\x_1,\ldots,\x_n,t) = \mean{\mathrm{e}^{- \Lambda_{n}(t)
    }}_{n}.
\label{Eq:SolGenPnlongtime}
\end{equation}
Integrating over all spatial variables, for $n\ge2$, we obtain
\begin{eqnarray}
 {\G_{n}}(t) 
=\mean{\F_{n}^0 \, \mathrm{e}^{-
     \mu\sum\limits_{i<j}^{n}\int_0^t\theta(a-|\Y_i(s)-\Y_j(s)|)\,\d s}}
 \label{Eq:AsymPnIntegrated} 
\end{eqnarray}
where now the average is over all possible $n$-uplets of tracers.
Finally, averaging the factorial moments over the realizations of
the velocity field,  we can define the effective reaction rate as
\begin{equation}
  \gamma_n=-\lim_{t\to\infty}\frac{1}{t}
  \ln{\frac{\overline{{\G}_n(t)}}{{\overline{\G_n(0)}}}}.
  \label{Eq:defrate}
\end{equation}
Note that the average over different realizations of the carrier flow,
denoted with an overline, is performed before taking the
logarithm. This important point will be further discussed in
Sect.~\ref{subsec:effective}.

\begin{figure}
  \centering
  \includegraphics[width=0.8\textwidth]{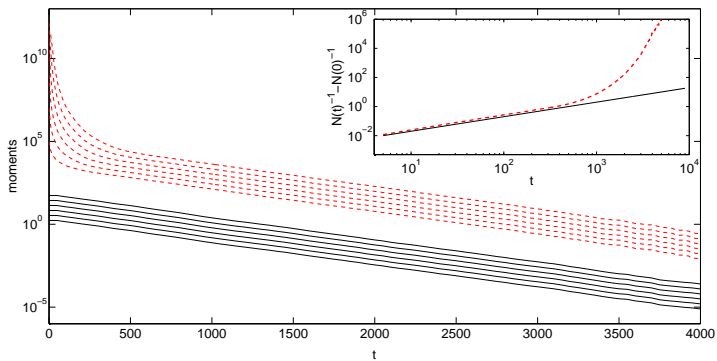}
  \caption{(Color online) Temporal evolution of $\langle N^p \rangle$
    for $p=1,2,3,4$ and initial values $N_0=2$ (black solid lines) and
    $N_0=100$ (red dashed lines). Simulations are obtained by
    Monte-Carlo simulations of the reactions using a delta-correlated
    in time velocity field (the so called Kraichnan flow, see
    Sec.~\ref{sec:dilute_kraich} for settings). Average is taken over
    reactions, realizations of the flow and initial positions of
    particles ($N_0=2$ and $N_0=100$ runs averaged over more than
    $10^7$ and $10^6$ realizations respectively). The parameters of
    the simulation are $\mu=1$, $\wp=0.1$ and $a/L=0.015$. The red
    dashed lines have been arbitrarily multiplied by a factor $1000$
    for the sake of representation. Inset: temporal evolution of the
    inverse of the average number of particles for $N_0$=100.  It
    displays the mean-field prediction $\langle N(t) \rangle_\mu
    \simeq N_0/(1+C\,N_0\,t)$ which is indeed expected to work at
    short times. \label{fig:mom_fn_time}}
\end{figure}

In the zero-dimensional limit, by definition the Heaviside function
inside equation \eqref{Eq:AsymPnIntegrated} is equal to one, and thus
the effective rate is simply given by $\gamma_n=\mu\,{n(n-1)}/{2}$, as
confirmed by numerical data (not shown). When the system is not
well-mixed at all scales, the effective rate \eqref{Eq:defrate} cannot
be computed for generic flows as multi-time and space correlations of
Lagrangian trajectories are involved. However, an important role
  is played by the case $n=2$ that only depends on the two-point
  motion. Indeed, it is easy to check from equation
  \eqref{Eq:AsymPnIntegrated} that $\gamma_n$ is an increasing
  function of $n$. In addition, the moments of the number of particles
  can be expressed as a linear combination of the factorial
  moments. The rate $\gamma_2$ is thus always the leading exponent and
  for all $p\ge1$ we have
\begin{equation}
  \langle N^p \rangle \sim \mathrm{e}^{-\gamma_2\,t}.
\end{equation}
This exponential decrease is expected to be valid at long times
regardless of the initial number $N_0$ of particles considered. This
behavior is evidenced in Fig.~\ref{fig:mom_fn_time} where the temporal
evolution of the moments $\langle N^p \rangle$ is displayed for
$N_0=2$, $100$ and different orders $p$. For $N_0=100$, the long-time
departure from the $\propto t^{-1}$ mean-field prediction is clear
from the inset.
Independently of the order of the moment considered, the long-time
dynamics depends only on the statistics of the relative distance of
one pair of tracers, which motivates the next section that is devoted
to the study of the two-point motion.

\section{Two-point motion}
\label{sec:2point}

\subsection{Effective rate}
\label{subsec:effective}

As the long-time behavior of the system is dominated by the
two-particle dynamics, it is natural to close the hierarchy
\eqref{Eq:DefMasterEq} at the second order and to look at the mean
numbers of particles and of pairs:
\begin{eqnarray}
  n_1(t) \equiv \overline{\mean{N(t)}}_\mu =
  \overline{\G_1(t)}&,\hspace{3mm}&
  n_2(t) \equiv\frac{1}{2}\,\overline{\mean{N(t)\,[N(t)-1]}}_\mu =
  \frac{1}{2}\overline{\G_2(t)}.
  \label{eq:def_n1n2}
\end{eqnarray}
The overbars designate the average with respect to the realizations of
the velocity field $\v$, which is assumed statistically homogeneous in
space so that the overbars include the average over the particle
initial positions.  Both quantities, $n_1(t)$ and $n_2(t)$, are
expected to exponentially decrease with the same rate $\gamma_2$,
which here and in the following will be denoted by $\gamma$.  An
effective equation describing such a behavior is simply given by
\begin{eqnarray}
  \frac{\d n_1}{\d t}  =  -\gamma\,n_2 &,\hspace{3mm}&  \frac{\d
    n_2}{\d t}  =  -\gamma\,n_2. \label{eq:effective_equation}
\end{eqnarray}
The effective rate $\gamma$ depends on the microscopic rate $\mu$, the
interaction radius $a$ and the statistics of the carrier flow through
equations \eqref{Eq:AsymPnIntegrated} and \eqref{Eq:defrate}. However,
these equations do not allow us to readily obtain an explicit
expression for $\gamma$. The dependence of such an expression on the
parameters of the system can be obtained by using standard tools of
dynamical system and statistical physics.

The effective rate $\gamma$, defined by equation \eqref{Eq:defrate},
depends on the particles trajectories only through $\F_2$ and thus is
determined by the two-particle statistics. Consequently, without loss
of generality, we will focus on a system initially having $N_0=2$
particles. For such a system, $\F_2^0 = 2$ and the average number of
pairs reads
\begin{equation}
  n_2(t)  = \overline{\mathrm{e}^{-\mu \int_0^t
      \theta(a-R(s))\,\d s} },
  \label{eq:n2}
\end{equation}
where $R(t)=|\R(t)| = |\Y_1(t)-\Y_2(t)|$ and the average is over all
realizations of the carrier velocity field $\v$.
The argument of the exponential in Eq.~\eqref{eq:n2} contains a time
integral that, for ergodic two-point dynamics, self-averages, i.e.
\begin{equation}
  \lim_{t\to\infty} \frac{1}{t} \int_0^t \theta(a-R(s))\,\d s =
  P_2^<(a) \equiv \mathbb{P}\{ R(s)<a \},
\end{equation}
where $\mathbb{P}$ is a shorthand notation for probability.  This
observation leads to the following naive prediction
\begin{equation}
  \gamma_{\rm naive} = \mu\,P_2^<(a).
  \label{eq:naive}
\end{equation}
This estimate is consistent with what could be expected on a heuristic
basis, however it is valid only when considering a fixed realization
of the carrier flow. This kind of setting corresponds to the so-called
\emph{quenched disorder} in statistical mechanics.  Indeed, let us fix
a realization $\v$ of the carrier flow and the initial positions
$\X_1(0)$ and $\X_2(0)$ of the pair of reacting particles. If we now
denote by $N(t)$ the variable which is equal to $2$ when the two
particles have not reacted and $0$ otherwise, we clearly have that
\begin{equation*}
\lim_{t\to\infty} \frac{1}{t} \ln \langle N(t)[N(t)-1] \rangle_\mu =
 - \lim_{t\to\infty} \frac{\mu}{t} \int_0^t
  \theta(a-R(s))\,\d s = -\mu \,P_2^<(a).
\end{equation*}
Notice that, here, the average is performed solely with respect to
reactions. The quenched effective rate is thus given by
\eqref{eq:naive}.  The average over fluid realizations can be
subsequently performed but it does not modify the result
\eqref{eq:naive} as $\ln \langle N(t)[N(t)-1] \rangle_\mu$ is a
self-averaging quantity.

The situation is quite different when considering an \emph{annealed
  disorder}, that is when the fluctuations of the carrier flow are
taken into account to compute the number of pairs, according to the
definition \eqref{eq:n2}. In this case, as $n_2$ is not a
self-averaging quantity, fluctuations of the carrier flow affect the
long-time behavior and, hence, $\gamma$ is not simply given by
\eqref{eq:naive}. In most physical situations, fluctuations of the
carrier flow and randomness of the initial particle positions cannot
be disregarded, so that \emph{annealed} averages must be performed to
define the effective rate \eqref{Eq:defrate}. As discussed in the
next subsections, in this case, the effective rate $\gamma$ can be
related to the large deviations of the fraction of time spent by two
particles within a distance less than $a$.

\subsection{Large deviations of the reaction time}
\label{subsec:largedev}

\begin{figure}[thb]
  \centering
    \includegraphics[width=0.49\textwidth]{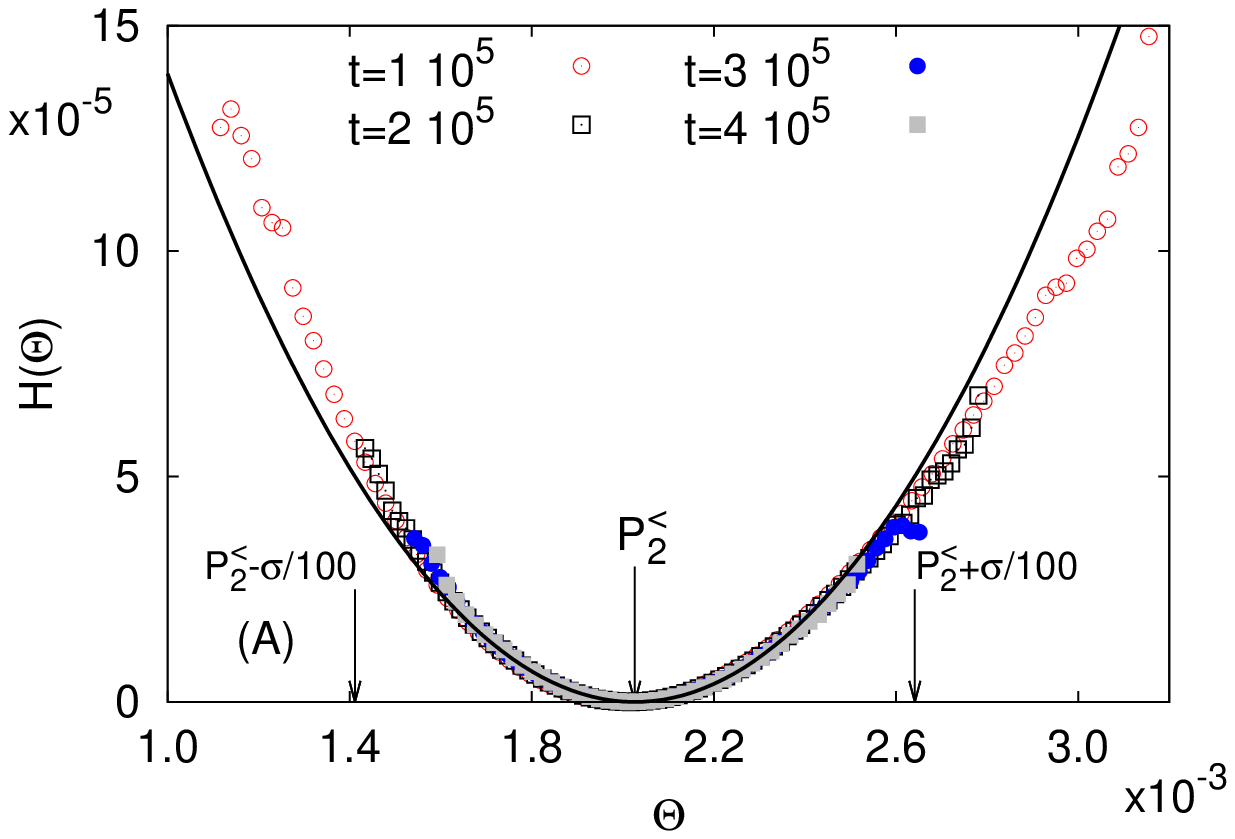}
  \includegraphics[width=.49\textwidth]{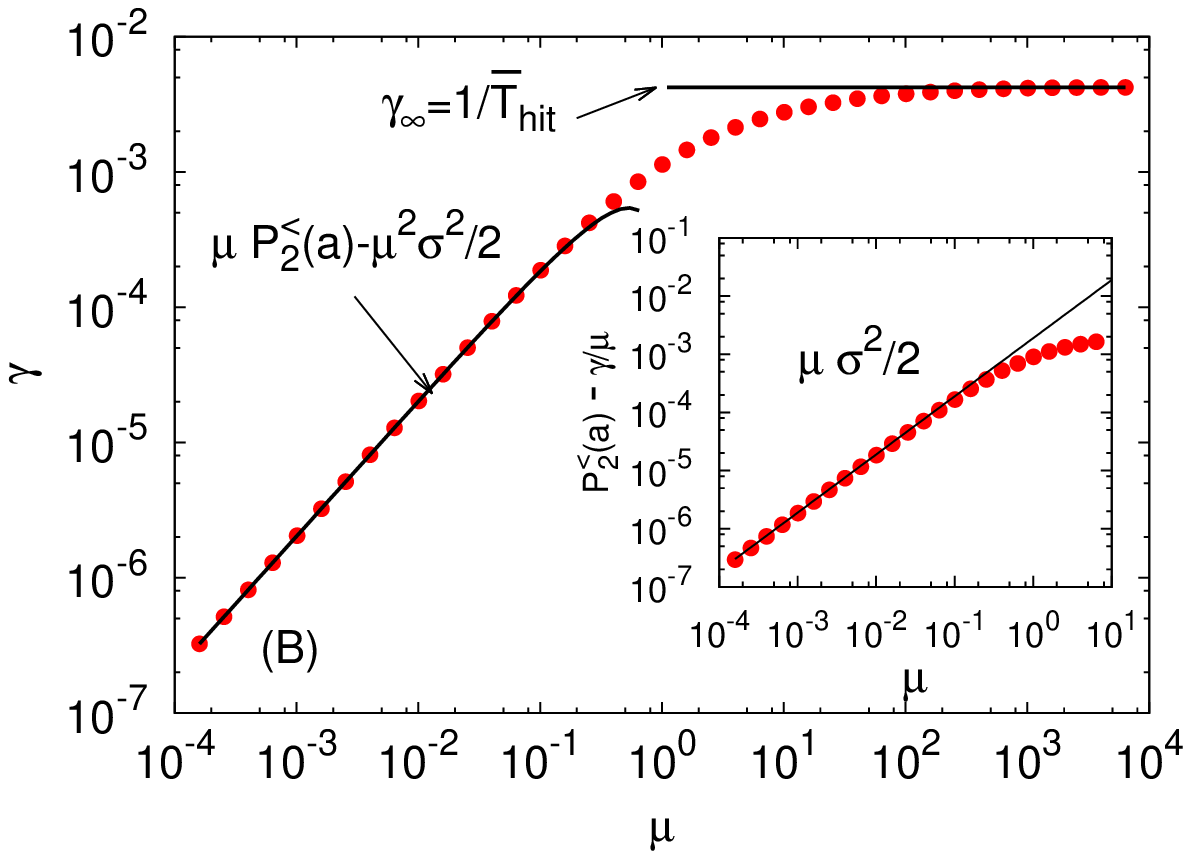}
  \caption{(Color online) Data obtained by using the Kraichnan flow,
    see Sec.~\ref{sec:dilute_kraich} for details. (A) The rate
    function $\mathcal{H}$ as a function of $\Theta$ for $a/L=5\cdot
    10^{-3}$ and $\wp=0.1$ at different times as labeled. The solid
    black line denotes the central limit theorem approximation
    $\mathcal{H}(\Theta)=(\Theta-P_2^{<}(a))^2/(2\sigma^2)$. The
    arrows show where the minimum $\mathcal{H}=0$ is attained, i.e.\
    for $\Theta=P_2^{<}(a)$, and the region of validity of the
    approximation. (B) $\gamma$ vs $\mu$ for $\wp=0.1$ and $a/L=5\cdot
    10^{-3}$ the solid lines shows the small (\ref{eq:quadratic_rate})
    and large (\ref{Eq:gammainfty}) $\mu$ asymptotic,
    respectively. The inset blows up the small $\mu$ asymptotics
    showing that $P_2^{<}(a)-\gamma/\mu\approx \mu \sigma^2/2$ so to
    make evident the quadratic CLT corrections as from
    Eq.~(\ref{eq:quadratic_rate}).}
  \label{fig:gamma_smallmu}
\end{figure}

Determining the long-time asymptotics of $n_2$ requires to understand
the behavior of the finite-time average
\begin{equation}
\Theta = \frac{1}{t} \int_0^t\theta(a-R(s))\,\d s.
\end{equation}
Under ergodicity hypotheses on the two-point dynamics, this
self-averaging quantity converges at long times to $P_2^<(a)$. When
the dynamics is sufficiently mixing, the deviations of $\Theta$ from
its average are described by a large-deviation
principle~\cite{Ellis85}. The probability density function (PDF)
$p(\Theta)$ thus has the asymptotic behavior
\begin{equation}
 - \lim_{t\to\infty} \frac{1}{t} \ln p(\Theta) = \mathcal{H}(\Theta),
 \label{Eq:DefRateFunction}
\end{equation}
where $\mathcal{H}$ is a convex positive rate function, which attains
its minimum, equal to zero, at $\Theta = P_2^<(a)$. Plugging this form
in \eqref{eq:n2} we obtain that
\begin{equation}
  n_2(t) \propto  \int_0^\infty \mathrm{e}^{-t\,(\mu\,\Theta +
    \,\mathcal{H}(\Theta))} \d \Theta.
\end{equation}
A saddle-point estimate of the integral shows that the effective rate
$\gamma=-\lim\limits_{t\to\infty} \frac{\ln n_2(t)}{t} $ is given by
the Legendre transform of the rate function $\mathcal{H}$:
\begin{equation}
  \gamma = \inf_{\Theta\ge0} [\mu\,\Theta + \mathcal{H}(\Theta)]\,.
  \label{eq:legendre}
\end{equation}
The convexity of $\mathcal{H}$ implies that $\gamma$ is a
non-decreasing function of $\mu$. For illustrative purposes, in figure
\ref{fig:gamma_smallmu} (A) we show the rate function for the Kraichnan
flow (a delta-correlated in time velocity field, see
Sec.~\ref{sec:dilute_kraich} for numerical settings and throughout
discussion of this kind of flows).

When the reaction rate $\mu$ is small compared to the scale of
variation of $\mathcal{H}$, the infimum in \eqref{eq:legendre} is
attained very close to the minimum of $\mathcal{H}$ at $\Theta =
P_2^<(a)$. There, as a result of the central limit theorem (CLT), the
rate function is approximately quadratic $\mathcal{H}(\Theta) \simeq
(\Theta-P_2^<(a))^2/(2\sigma^2)$ (represented by the black solid line
in figure \ref{fig:gamma_smallmu} (A)) and the infimum in
\eqref{eq:legendre} is attained for $\Theta = \Theta_\star = P^<_2(a)
- \mu\sigma^2$, so that
\begin{equation}
  \gamma \simeq \mu\,P_2^<(a) - \mu^2\, \frac{\sigma^2}{2}\,,
  \label{eq:quadratic_rate}
\end{equation}
where $\sigma^2/t$ is the variance of the fraction of
time $\Theta$ spent by the pair at a distance less than $a$.
This behavior is apparent figure \ref{fig:gamma_smallmu} (B) where the
effective rate is displayed for the same flow.
The expression \eqref{eq:quadratic_rate} is valid as long as
$\Theta_\star$ is close enough to $P_2^<(a)$ to neglect the
sub-leading terms in the quadratic approximation for $\mathcal{H}$,
i.e.\ in the domain of validity of the CLT. Notice that
\eqref{eq:quadratic_rate} requires $\mu\ll P_2^<(a)/\sigma^2$.
Observe also that the naive estimate \eqref{eq:naive} for a
\emph{quenched} velocity field only captures the linear behavior of
$\gamma$ close to $\mu=0$.

The asymptotic behavior of $\gamma$ at large values of $\mu$ relates
to that of the rate function $\mathcal{H}$ at small values of the
fraction of time $\Theta$ spent below $a$, indeed, from
\eqref{eq:legendre} we have
\begin{equation}
  \lim_{\mu\to\infty}\gamma=\lim_{\Theta\to0}\mathcal{H}(\Theta).
  \label{Eq:gammainf}
\end{equation}
The events leading to small $\Theta$'s correspond to those in which the
particle-pair separation remains larger than $a$ for a very long
time. More precisely, the value of $\mathcal{H}$ at $\Theta=0$ relates
to the distribution of the random time $T_\mathrm{hit}$ needed by
pairs that are initially far apart (e.g.\ at separations of the order
of the system size) to approach each other at a distance less than
$a$. We can indeed write
\begin{equation}
  \mathbb{P}(\Theta<\varepsilon) =
  \mathbb{P}(\Theta<\varepsilon\,|\,T_\mathrm{hit}>t) \,
  \mathbb{P}(T_\mathrm{hit}>t)
  + \mathbb{P}(\Theta<\varepsilon\,|\,T_\mathrm{hit}<t) \,
  \mathbb{P}(T_\mathrm{hit}<t).
\end{equation}
When $T_\mathrm{hit}>t $, the two particles have never been at a
distance less than $a$, which implies
$\mathbb{P}(\Theta<\varepsilon\,|\,T_\mathrm{hit}>t) =1$ for any
$\varepsilon>0$. Conversely, when $T_\mathrm{hit}<t$, the fraction of
time $\Theta$ is finite and
$\mathbb{P}(\Theta<\varepsilon\,|\,T_\mathrm{hit}<t)\to 0$ for
$\epsilon\to0$. Hence, it follows that
\begin{equation}
  \mathcal{H}(0) = - \lim_{t\to\infty}
  \frac{1}{t} \ln \mathbb{P}(T_\mathrm{hit}>t).\label{Eq:H0Thit}
\end{equation}
Different behaviors of the effective rate are thus expected depending
on the tail of the distribution of $T_\mathrm{hit}$. For the proposed
large-deviation approach to be valid, one expects
$\mathbb{P}(T_\mathrm{hit}>t)$ to decrease at least as fast as an
exponential. Otherwise, one would obtain $\mathcal{H}(0)=0$, 
violating the convexity of $\mathcal{H}$.  When
$\mathbb{P}(T_\mathrm{hit}>t)$ decreases faster than an exponential,
$\mathcal{H}(0)= \infty$, and the effective rate $\gamma\to\infty$ for
$\mu\to\infty$. The way it diverges cannot be obtained from the
distribution of $T_\mathrm{hit}$ only, as it depends on the functional
form of $\mathcal{H}$ close to $\Theta=0$.
In the intermediate case, when $\mathbb{P}(T_\mathrm{hit}>t)$
decreases as an exponential, $\mathcal{H}(0)$ is finite and the
effective rate approaches a finite value $\gamma_\infty =
\mathcal{H}(0)$ when $\mu\to\infty$. Depending on whether $\mu^\star =
-\d \mathcal{H}/\d \Theta|_{\Theta=0}$ is finite or not, $\gamma$
saturates to $\gamma_\infty$ for $\mu>\mu^\star$ or approaches it
asymptotically. Note that an exponential behavior for the tail of the
distribution of $T_\mathrm{hit}$ is not a particular case but is
actually expected to be a quite generic circumstance. Indeed, when
considering times much longer than all relevant timescales of the
flow, memory loss implies that the hittings of $a$ form a Poisson
process, and the hitting time is an exponential random variable. This
is directly corroborated when considering the Kraichnan flow (data not
shown). For an exponential distribution of the hitting time we simply
obtain
\begin{equation}
\gamma_\infty=\frac{1}{\overline{T_{\rm hit}}}.\label{Eq:gammainfty}
\end{equation}
This asymptotic behavior is also observed in figure
\ref{fig:gamma_smallmu} (B) for large values of $\mu$.

In the next subsection we introduce and study a phenomenological model
for which the rate function can be explicitly computed.

\subsection{A phenomenological model for reactions in bounded chaotic
  flows}
\label{subsec:heuristic}

\begin{figure}
  \centering
  \includegraphics[width=\textwidth]{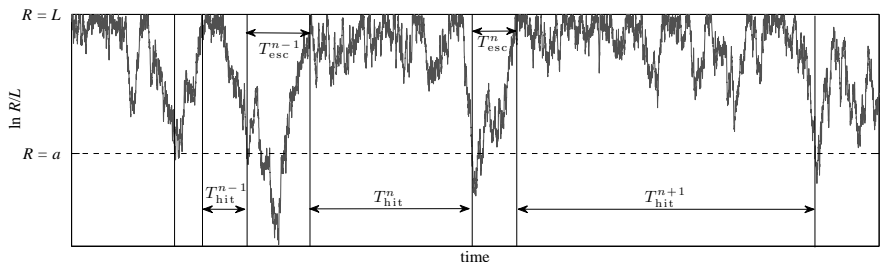}
  \caption{Illustration of the decomposition in hitting events of
    duration $T_{\rm hit}^n$ and escapes of duration $T_{\rm esc}^n$
    as discussed in the text.}
  \label{fig:sketch_exit}
\end{figure}

The essence of chaotic flows is the competition between stretching and
folding. Stretching results from the fact that close trajectories
separate exponentially at a rate given by the largest Lyapunov
exponent $\lambda$. Folding, which is generally due to boundary
conditions, prevents the inter-particle distance to grow indefinitely
and is necessary for the system to converge to a statistical steady
state. Qualitatively, the time evolution of the distance $R$ between
two particles resembles the trajectory shown in
Fig.~\ref{fig:sketch_exit}. The particle pair spends long times at a
distance of the order of the size $L$ of the domain and performs rare
excursions to very small separations.

When interested in particles reacting within a distance $a\ll L$, it
is quite natural to decompose the inter-particle dynamics in a
sequence of consecutive phases, each corresponding to a given
approaching event and separated by returns to $L$.  Starting from
$R=L$, a first time subinterval is required to hit $a$ for the first
time. Then the particle separation grows to touch again $L$. The
time interval $[0,t]$ is thus decomposed in $n_{\rm rea}$ random
subintervals of length $T^n_{\rm hit}+T^n_{\rm esc}$, where $T^n_{\rm
  hit}$ is the hitting time from $R=L$ to $R=a$ and $T^n_{\rm esc}$
the time needed to reach $L$ for the first time starting from $a$ (see
figure \ref{fig:sketch_exit}). Because of the memory loss
  occurring when the pairs are at a distance $L$, these sub-intervals
  can be considered independent and identically distributed. The
first substage corresponds to many recurrences to $R=L$ during time
lengths that are independent from each other because of the imposed
boundary condition. The hitting time is dominated by the sum of these
recurrence times and this leads to assume that $T^n_{\rm hit}$ has an
exponential distribution of mean $\overline{T_{\rm hit}}$. Concerning
the escape time $T_{\mathrm{esc}}^n$, it is clear that when $a\ll L$,
its behavior is dominated by the long-time convergence to the Lyapunov
exponent. We can thus assume that it is deterministic, namely
$T^n_{\rm esc} \approx T_{\rm esc} = (1/\lambda)\ln (L/a)$. In
addition, in this limit, we always have $T_{\rm esc} \ll T^n_{\rm
  hit}$ and thus
\begin{equation}
  t \approx \sum_{n=1}^{n_{\rm rea}}\left [T^n_{\rm hit} +T^n_{\rm esc}
  \right] \approx \sum_{n=1}^{n_{\rm rea}}T^n_{\rm hit}.
\end{equation}
According to the discussion of the previous subsection, the times
$T^n_{\rm hit}$ are exponentially distributed and $n_{\rm rea}$ is a
Poisson random variable of mean $t/\overline{T_{\rm hit}}$.

During the $n$-th interval of length $T_{\rm esc}^n$, the particles
spend a time $\alpha_n\,T_{\rm esc}^n$ below a distance $a$. The total fraction
of time spent at $R<a$ can thus be written as
\begin{eqnarray}
  \Theta=\frac{1}{t}\sum_{n=1}^{n_{\rm rea}} \alpha_n\,T^n_{\rm esc}
  \approx \frac{n_{\rm rea}}{t}\,\overline{\alpha}\,T_{\rm esc},
  \label{Eq:timedecomp}
\end{eqnarray}
where $\overline{\alpha}$ designates the average fraction of time
spent below $a$ during $T_{\rm esc}$.  We have assumed in this model
that the fluctuations of $\alpha$ do not play an important
role. Notice that under these assumptions, $P_2^<(a) =
\overline{\Theta} = \overline{\alpha}\, T_{\rm esc} /\overline{T_{\rm
    hit}}$.  We see from equation~\eqref{Eq:timedecomp} that $\Theta$
is proportional to $n_{\rm rea}$. Its probability density is thus
given by the Poisson distribution. Using the Stirling formula when
$t\to\infty$, we obtain
\begin{equation}
  \mathcal{H}(\Theta)=\frac{1}{\overline{T_{\rm hit}}}   
  -\frac{\Theta}{\overline{\alpha} \,T_{\rm esc}}\left (1+\ln{\left[
        \frac{\overline{\alpha}\,
          T_{\rm esc}}{\Theta\, \overline{T_{\rm hit}} }
      \right]}    \right).\label{Eq:RateFuncHeur}
\end{equation}
It is easy to check that $\mathcal{H}(\Theta)$ given by
\eqref{Eq:RateFuncHeur} is a convex function that vanishes at its
minimum reached at $\Theta_{\rm min} =\overline{\alpha}\, T_{\rm esc}
/\overline{T_{\rm hit}}=P_2^<(a)$, as expected. Also, one sees that $
\mathcal{H}(0)={1}/{\overline{T_{\rm hit}}}$ and
$\mathcal{H}^\prime(0)=-\infty$, where the prime denotes the
derivative with respect to $\Theta$.  In the light of the
considerations of the previous section, this implies that the
effective reaction rate $\gamma$ asymptotically converges to
$\gamma_{\infty} = 1/\overline{T_{\rm hit}}$ when $\mu\to\infty$. With
the explicit form \eqref{Eq:RateFuncHeur} for the rate function we
actually obtain
 \begin{equation}
   \gamma = \frac{1}{\overline{T_{\rm hit}}}\left[1-{\rm e}^{-P_2^<(a)\,\overline{T_{\rm hit}}\,\mu}\right].\label{Eq:gammaModel}
\end{equation}
Note that this expression also reproduces the small-$\mu$ asymptotics
\eqref{eq:quadratic_rate} with $\sigma^2=P_2^<(a) ^2 \overline{T_{\rm
    hit}}$.

The explicit form \eqref{Eq:gammaModel} relies on the strong
simplifications of the particle dynamics that we have made
(\textit{cf.} Fig.~\ref{fig:sketch_exit}).  However, this model is
particularly useful to understand how $\gamma$ depends on the reaction
radius $a$, the flow compressibility, and its chaoticity that will
affect both $P_2^<$ and $\overline{T_{\rm hit}}$.  The time
$\overline{T_{\rm hit}}+T_{\rm esc} \approx \overline{T_{\rm hit}}$
can be seen as the recurrence time from $a$ to $a$. Therefore, by the
\emph{Kac Recurrence Lemma} \cite{KacLemma}, it is expected that
$\overline{T_{\rm hit}}\sim 1/P_2^<(a) $. However, the proportionality
factor depends on the Lyapunov exponent and on the compressibility,
and thus on the detailed properties of the carrier flow.

The model presented in this section, although phenomenological, is
based on quite general assumptions, such as the existence of a finite
correlation time and of at least one positive Lyapunov exponent.  The
precise shape of the rate function \eqref{Eq:RateFuncHeur} can vary
from one system to another when more realistic dynamics is considered.
It is however reasonable to expect that the general qualitative
properties of the effective rate $\gamma$ remain essentially the same.

\section{Applications to very dilute suspensions in compressible
  Kraichnan flows}
\label{sec:dilute_kraich}

To illustrate and further validate the approach developed in the
previous section, we focus here on two-particle motion in a special
class of random velocity fields, that is the smooth (compressible)
Kraichnan ensemble \cite{gawedzki2000phase,FalkovichG.2001}. The main
advantage of such a class of stochastic flows is that, thanks to time
uncorrelation, we have analytical control on most of the quantities we
need. Further it is very efficient to implement numerically so to gain
high quality statistics.

\subsection{The Kraichnan velocity ensemble}
We focus on dimensions $d>1$. The velocity $\v$ is a homogeneous,
isotropic, Gaussian spatially smooth field, with zero mean and
two-point correlation function
\begin{equation}
\overline{v_i(\vec{0},t)v_j(\vec{r},t')}=2D_{ij}(\vec{r})\delta(t-t')\,,
\qquad D_{ij}(\vec{r})=D_0 \delta_{ij}-d_{ij}(\vec{r})/2\,.
\label{eq:corrKraich1}
\end{equation}
For small separations $r=|\vec{r}|\to 0$ the spatial part of the
correlation is prescribed according to \cite{gawedzki2000phase}
\begin{equation}
  d_{ij}(\vec{r})=D_1  [(d+1-2\wp)\delta_{ij} r^2 +2(\wp d-1)r_ir_j]\,.
\label{eq:corrKraich2}
\end{equation}
The flow parameters are as follows: the diffusivity $D_0$ controls the
single particle dispersion properties; $D_1$ has the dimension of an
inverse time and sets the intensity of velocity gradients;
finally, $\wp \in [0:1]$ controls the compressibility degree as
defined in Sect~\ref{ssec:settings}, its extreme values correspond to
incompressible and potential velocity fields, respectively.  When
advecting particles with the velocity field defined by
Eqs.~(\ref{eq:corrKraich1})-(\ref{eq:corrKraich2}), we will always
consider the presence of boundary conditions, which are necessary in
order to ensure statistical steady state properties (see also below the
description of the numerical implementation).

As for the numerical implementation we consider two possible schemes.
The first, which is useful when considering many particles (this is
that used for Fig.~\ref{fig:mom_fn_time}), consists in generating a
random velocity field, with periodic boundary conditions, as the sum
of independent Fourier modes with $|\vec{k}|\le 2$ and amplitudes such
that Eqs.\eqref{eq:corrKraich1}-\eqref{eq:corrKraich2} are satisfied
at small scales. When interested in two-particle motion, we use a
second scheme, which is much more efficient. It evolves directly the
stochastic equation for the separation $\R$ between the particles. We
use a discretization by means of a standard Euler-Ito scheme that
reads
\begin{eqnarray}
  &&R_i(t+\mathrm{d}t)=R_i(t)+\sqrt{2\mathrm{d}t}V_i(\vec{R},t)
  \quad\mbox{ for } i=x,y  \nonumber\\
 &&\mbox{with } V_x=\sqrt{D_1(1+2\wp)}R_x\eta_x - \sqrt{D_1(3-2\wp)}
 R_y\eta_y\nonumber\\
  &&\phantom{\mbox{with }} V_y=\sqrt{D_1(1+2\wp)}R_y\eta_x +
  \sqrt{D_1(3-2\wp)} R_x\eta_y\nonumber\,,
\end{eqnarray}
which only requires to generate two independent zero-mean Gaussian
random variables, $\vec{\eta}_i$. The above equations indeed prescribe
that $\overline{V_i(\vec{R},t) V_j(\vec{R},t)}= d_{ij}(\vec{R})$, with
$d_{ij}$ given by Eq.(\ref{eq:corrKraich2}) for $d=2$. Finally, to
ensure a statistically steady dynamics, reflective boundary conditions
are imposed at $|R_i|=L/2$. In principle, this scheme can be
generalized also to more than two particles \cite{FMNV99}.

Particles evolving in such flows with the dynamics
Eq.~(\ref{eq:tracers}) define a stochastic dynamical systems of which
we know explicitly the Lyapunov exponents,
$\lambda_k=D_1[d(d-2k+1)-2\wp(d+(d-2)k)]$ ($k=1,\ldots,d$), and the
rate function of the stretching rates, which is quadratic (see the
review \cite{FalkovichG.2001} for a summary of the known properties of
Kraichnan flows). In particular, the particle separation $r=R(t)$
follows a diffusion process whose generator, restricted to homogeneous
and isotropic sectors, is simply given by
\begin{equation}
  M_2=\frac{D_1(d-1)(1+2\wp)}{r^{\mathcal{D}_2-1}}\frac{\partial}{\partial
    r}\left(r^{\mathcal{D}_2+1} \frac{\partial}{\partial
      r}\right)\label{Eq:defM2},
\end{equation}
where
\begin{equation}
\mathcal{D}_2 =  \frac{d-4\wp}{1+2\wp}= d-\wp \frac{2(d+2)}{1+2\wp}\,.
\label{eq:d2kraich}
\end{equation}
To ensure the convergence to a statistically stationary regime, the
particle dynamics and the generator $M_2$ must be defined on a compact
set; this is done by supplying boundary conditions at the domain
border. The resulting stationary probability $P_2^<(a)$ of finding two
particles at a distance less than $a$ is
\begin{equation}
P_2^<(a)\sim (a/L)^{\mathcal{D}_2}\,\, \,\,\,\,\, \mbox{for  }
\,\,\,\,\,a\ll L\label{Eq:P2_D2}
\end{equation}
meaning that $\mathcal{D}_2$ defined in (\ref{eq:d2kraich}) is the
correlation dimension.

It is worth noticing that $\mathcal{D}_2=d$ for $\wp=0$, which simply
means volume preserving dynamics in the incompressible limit, and that
$\mathcal{D}_2=0$ for $\wp_c =d/4$, which is the value of the
compressibility $\wp$ above which the first Lyapunov exponent becomes
negative. The latter property indicates that above a critical
compressibility $\wp_c$ there is a transition to a regime of particles
trapping with particles asymptotically collapsing onto a single point
see, e.g., Ref.~\cite{gawedzki2000phase}. Notice that this regime
cannot be realized for $d>4$.

\subsection{Feynman--Kac approach}

We now turn to a direct computation of the effective rate $\gamma$ by
using the generator \eqref{Eq:defM2} of the diffusion process
$R(t)$. Note that by performing the change of variable
$x=\log{(r/L)}$, $M_2$ reduces to
 \begin{equation}
\tilde{M}_2=\lambda_1\frac{\partial}{\partial
  x}+\frac{\Delta}{2}\frac{\partial^2}{\partial^2
  x}\label{Eq:defM2log},
\end{equation}
where $\lambda_1=D_1(d-1)(d-4\wp)$ is the largest Lyapunov exponent
and $\Delta=2D_1(d-1)(1+2\wp)$ is the variance of its finite-time
fluctuations. Hence, for the Kraichnan model $\ln [R(t)]$ is a
Brownian motion with drift $\lambda_1$ and diffusion coefficient
$\Delta/2$ \cite{gawkedzki2004sticky}. Note also that the correlation
dimension is expressed as
\begin{equation}
\mathcal{D}_2=2\lambda_1/\Delta\,.
\label{eq:d2lambdadelta}
\end{equation}

To obtain the effective rate we need to compute the large-time
dynamics of the number of pair $n_2(t)$ given by equation
\eqref{eq:n2}. Conditioning on the initial separation between the
particles, we define
 \begin{equation}
\psi(x,t)=\overline{\left[\left.e^{-\mu \int_0^t
        \theta(\log{(a/L)}-x_s)ds}\right| x{_{s=0}=x}\right]},
\end{equation}
where $x_s$ is the Brownian motion with drift defined by the generator
$\eqref{Eq:defM2log}$. By definition $x_s$ is restricted to the
half-plane $x<0$ and we assume reflecting boundary conditions at $x=0$
(corresponding to the domain boundary $r=L$). The effective
rate is thus given by the decay rate of $\psi(x,t)$. The Feynman-Kac
formula \cite{donsker1975asymptotic,oksendal1998stochastic}
states that $\psi$ is the solution of the partial differential
equation
  \begin{equation}
    \frac{\partial\psi}{\partial
      t}=\tilde{M}_2\psi-\mu\theta(\log{(a/L)}-x)\psi,\hspace{2mm}\psi(x,0)=1,\hspace{1mm}
    \psi|_{x=-\infty}=0,
    \hspace{1mm}\left.\frac{\partial\psi}{\partial
        x}\right|_{x=0}=0. \label{Eq:FeynmanKac}
\end{equation}
The boundary condition at $x=0$ comes from the no-flux condition
expressed for the backward operator of the reflected Brownian motion
with drift defined by \eqref{Eq:defM2log} (see
e.g. \cite{Gardiner_1985}). The effective rate $\gamma$ is thus given
by the modulus of the largest eigenvalue of the operator
$\mathcal{L}=\tilde{M}_2-\mu\theta(\log{(a/L)}-x)$. The
eigenvalue problem $\mathcal{L}\psi_s=s\psi_s$ can be easily
solved finding the solutions for $x<\log{(a/L)}$ and $x>\log{(a/L)}$
and then imposing that $\psi_s$ and its derivative are continuous at
$x=\log{(a/L)}$. Finally, imposing the boundary condition at $x=0$
gives the following transcendental equation for the eigenvalues $s$:
\begin{eqnarray}
  -\frac{\left(\lambda _1^2-\lambda _1 \sqrt{\lambda _1^2+2 \Delta  (\mu
        +s)}+2 \Delta  s\right)}{\sqrt{\lambda_1^2+2 \Delta  s}}  \sinh
  \left(\frac{\log (a/L) \sqrt{\lambda _1^2+2 \Delta  s}}{\Delta
    }\right)+\hspace{2mm}&&\hspace{11mm} \label{Eq:EigenValues}\\   %
  \left(\sqrt{\lambda _1^2+2 \Delta  (\mu +s)}-\lambda _1\right) \cosh
  \left(\frac{\log (a/L) \sqrt{\lambda _1^2+2 \Delta  s}}{\Delta
    }\right)=0. \nonumber
\end{eqnarray}
All solutions of equation \eqref{Eq:EigenValues} are negative and the
absolute value of the largest one corresponds to the effective rate
$\gamma$. Taking the limit $\mu\to\infty$ in equation
\eqref{Eq:EigenValues} leads to a transcendental equation for the mean
value of the hitting time $T_{\rm hit}$. In addition, solving
\eqref{Eq:EigenValues} for small $\mu$ and using the asymptotic value
of $\gamma$ (\ref{eq:quadratic_rate}) directly gives $P_2^<(a)$ and
the variance $\sigma^2/t$ of the fraction of time $\Theta$ spent by a
pair of particles at a distance less than $a$:
\begin{eqnarray}
  &&P_2^<(a)= \left(\frac{a}{L}\right)^{\mathcal{D}_2}\\
  &&\sigma^2=\frac{4 \,P_2^<(a)  }{\mathcal{D}_2\lambda_1}
  \left[1- P_2^<(a)  \left(    1-\log\left[   P_2^<(a)  \right] \right
    )   \right]. \label{eq:sigmaKraichnan}
\end{eqnarray}

\subsection{Effective rate for the Kraichnan flow}

The Feynman-Kac calculations presented in the previous section are
independent of the space dimension $d$. This is a direct consequence
of the fact that the diffusive behavior of the logarithm of the
relative distance $x=\log{(r/L)}$ does not present any dependence on
$d$ other than through $\lambda_1$ and $\Delta$ (see equation
\eqref{eq:d2lambdadelta}). In the following, we will thus focus on the
two-dimensional case, which is the simplest non-trivial example. In
particular, we shall consider compressibility values $\wp<\wp_c=1/2$
to stay away from the trapping regime, which is trivial from the point
of view of reaction dynamics.

As discussed at length in the previous sections, the long-time
statistics of very dilute systems is dominated by the two-particle
effective reaction rate $\gamma$. This is confirmed by numerical
simulations of the compressible Kraichnan flow: as shown in
Fig.~\ref{fig:mom_fn_time} the decay of the moments of the number of
particles is indeed asymptotically dominated by
$\gamma$. Therefore, here, we focus on the two-point dynamics.

\begin{figure}[b!]
  \centering
  \includegraphics[width=.95\textwidth]{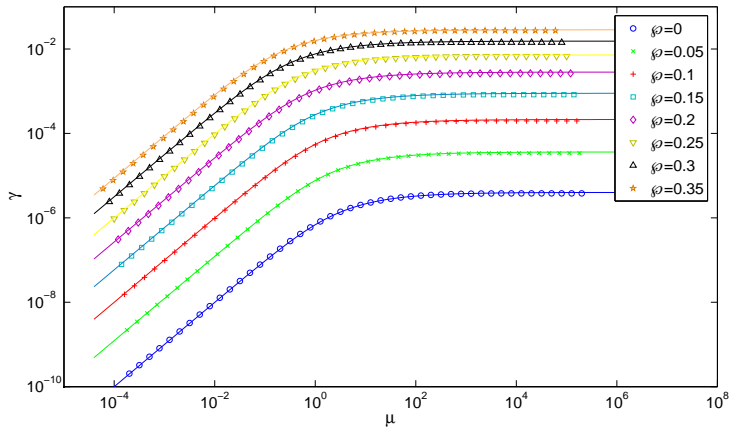}
  \caption{(Color online) $\gamma$ vs $\mu$ for different values of
    $\wp$. Solid lines are obtained by the Feynman-Kac calculations
    \eqref{Eq:EigenValues} and symbols by numerical simulations. The
    Lyapunov exponent $\lambda_1$ decreases from $2$ to $0.6$ when the
    compressibility is increased from $0$ to $0.35$ respectively.}
  \label{fig:gamma_largemu}
\end{figure}

The effective rate as a function of $\mu$ for different values of the
compressibility is displayed in figure \ref{fig:gamma_largemu}. Both,
Feynman-Kac calculations and numerics are in good agreement. Tiny
discrepancies are due to the lack of numerical precision when
estimating the exponential in \eqref{eq:n2} for large $\mu$.
It is apparent in figure \ref{fig:gamma_largemu} that compressibility
enhances reactions. The same behavior is also observed in $d=3$.

\begin{figure}[h!]
  \centering
  \includegraphics[width=.49\textwidth]{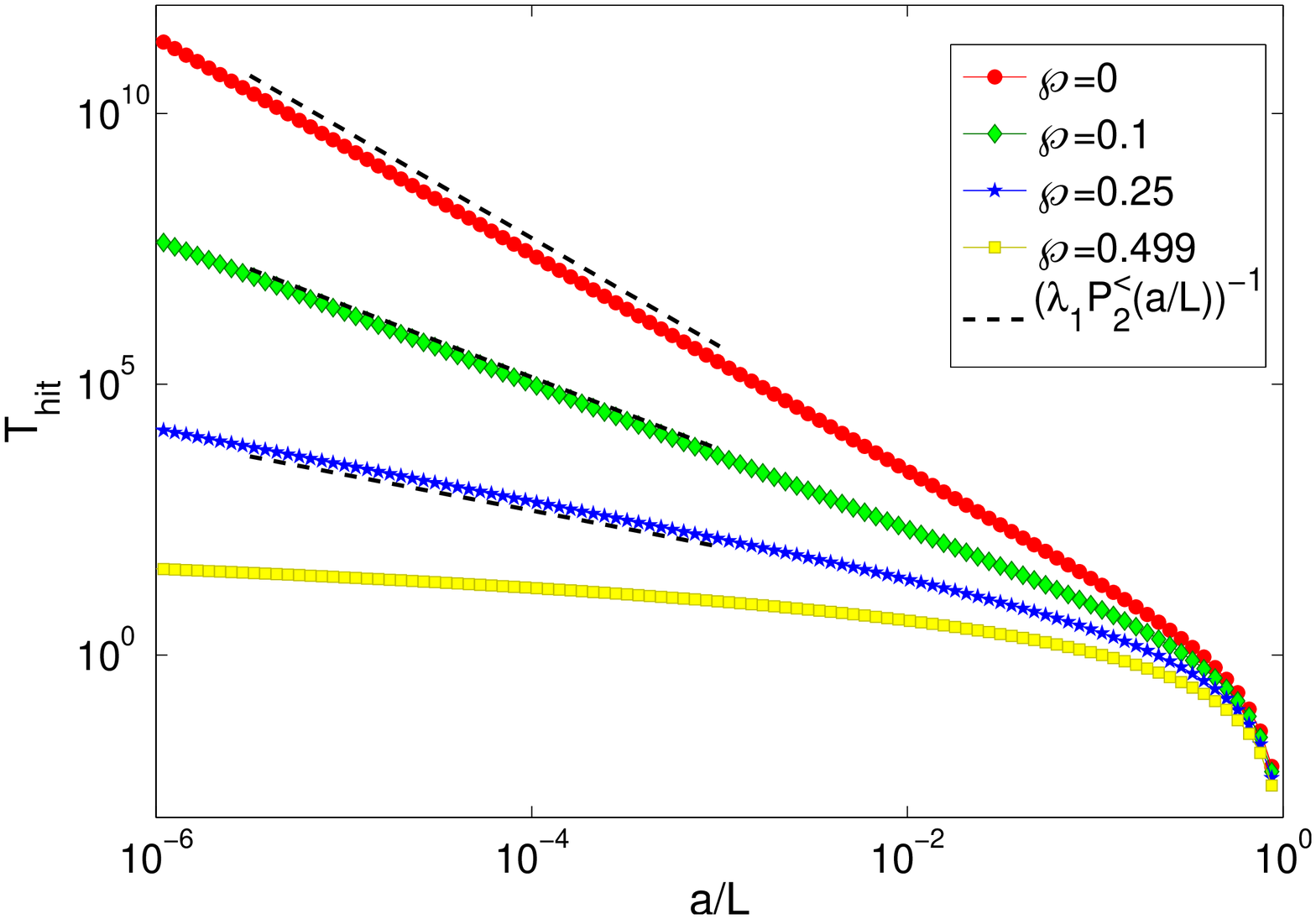}
  \includegraphics[width=.49\textwidth]{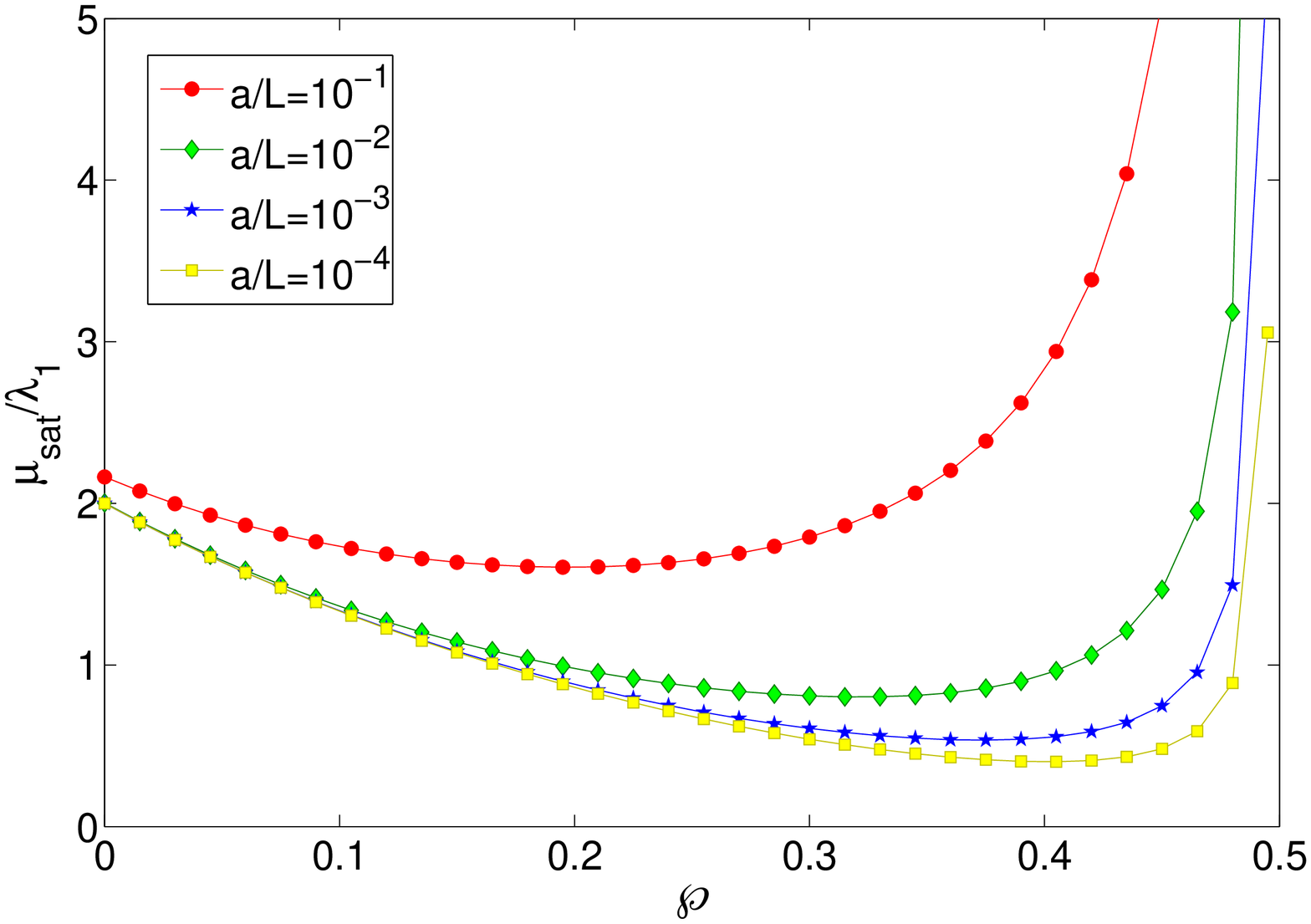}
  \caption{(Color online) (A) $\overline{T_{\rm hit}}$ as function of $a/L$ for
    different values of the compressibility. The black dashed line
    correspond to $(\lambda_1 P_2^<(a/L)$ for each compressibility.
    (B) Dependence of $\mu_{\rm sat}/\lambda_1$ on the compressibility
    $\wp$ for different values of $a/L$. For both figures
    $\overline{T_{\rm hit}}$ was directly obtained from
    \eqref{Eq:EigenValues} in the limit $\mu\to\infty$ .}
  \label{fig:mu_sta}
\end{figure}

The effective rate clearly exhibits two different regimes for small
and large values of $\mu$. For small $\mu$, the observed linear
behavior is given by the naive approximation $\gamma_{\rm
  naive}\approx \mu P_2^>(a)$ (see equation \eqref{eq:naive}). This
approximation remains valid as long as $\mu$ is much smaller than the
fluctuations of the fraction of time $\Theta$. From the central limit
theorem prediction \eqref{eq:quadratic_rate}, we infer that this
crossover takes place for $\mu^*\sim P_2^>(a)/\sigma^2$. Using
\eqref{eq:sigmaKraichnan} we thus obtain that the linear behavior on
the microscopic rate, only holds for $\mu\ll\mathcal{D}_2 \lambda_1$,
as confirmed in figure \ref{fig:gamma_largemu}.  On the other hand,
for large $\mu$, the rates saturates to the value $1/\overline{T_{\rm
    hit}}$. We can estimate the transition to the saturated regime by
extrapolating the naive approximation to the saturated value:
$\mu_{\rm sat}\,P_2^<(a)=1/\overline{T_{\rm hit}}$. As stated at the
end of section \ref{subsec:heuristic}, by the \emph{Kac Recurrence
  Lemma} \cite{KacLemma}, it is expected that $\overline{T_{\rm
    hit}}\sim (1/\lambda_1)P_2^<(a) $.
This is confirmed in figure \ref{fig:mu_sta} (A), which shows
$\overline{T_{\rm hit}}$ as function of $a/L$ for various values of
the compressibility.
We thus have that $\mu_{\rm sat}\sim\lambda_1$. The proportionality
factor depends in a non trivial way on the correlation dimension
$\mathcal{D}_2$ and $a/L$. Figure \ref{fig:mu_sta} (B) displays
$\mu_{\rm sat}/\lambda_1$ as a function of the compressibility $\wp$
for different values of $a/L$.  In the limit $\wp\to \wp_c$, $\mu_{\rm
  sat}$ has a finite value while $\lambda_1=0$ and thus $\mu_{\rm
  sat}/\lambda_1$ diverges.  Notice that in such limit $\Delta \to
3D_1(d-1)d$ while $\lambda_1\to 0$, meaning that (\ref{Eq:defM2log})
reduces to the diffusion operator without drift.  For larger values of
the compressibility all Lyapunov exponents become negative and the
particles concentrate on points.

\subsection{Beyond the Kraichnan model}
We conclude this section stressing that the Kraichnan model, for which
the problem of determining the effective particle-pair reaction rate
can be analytically solved, can also be seen as a first approximation
for more general (time-correlated) flows. Indeed the essence of
(\ref{Eq:defM2log}) is that the separation $R(t)$ between two
particles pairs which start at time $0$ with a separation $R_0$
follows the lognormal distribution
\begin{equation}
  P(R,t) = \frac{1}{R}\frac{1}{\sqrt{2\pi\Delta}} \exp\left[
    -\frac{[\ln(R/R_0)-\lambda_1 t]^2}{2\Delta t}\right].
  \label{eq:lognormal}
\end{equation}
This law, which is exact in the Kraichnan model
\cite{FalkovichG.2001}, represents a fairly good approximation for
generic chaotic systems provided fluctuations are not too wild, see
e.g. Ref.\cite{paladin1987anomalous,Crisanti93}. Therefore, within the
limit of validity of the lognormal approximation for generic chaotic
flows we can assume that the effective rate can be reasonably
approximated assuming the validity of (\ref{Eq:defM2log}) with
$\lambda_1$ and $\Delta$ depending on the flow details.  Furthermore,
for a smooth velocity field, if the dynamics is sufficiently mixing,
statistically reversible and stationary, there is a general result
stating that the correlation dimension is exactly expressed by
Eq.~(\ref{eq:d2lambdadelta})
\cite{grassberger1988,Bec2004}. Therefore, we can conjecture that
whenever deviations from Eq.~(\ref{eq:d2lambdadelta}) are small the
effective rate computed as in the previous section with suitable
values of $\lambda_1$ and $\Delta$ should provide a reasonably good
approximation for the effective reaction rate.

\section{Conclusion}
\label{sec:conclusion}

In this work we have considered a dilute system of particles
transported by a compressible flow.  The dynamics of particles was
supplemented by the binary reaction $A+A\rightarrow\varnothing$ with a
rate $\mu$. Particles can react only when their separation is below a
finite interaction radius $a$. We focused on a situation where
transport dominates over diffusion (high Schmidt numbers) and thus
particles are only driven by the carrier flow. The interest of the
setting of this work is that, independently of the initial number of
particles, the standard mean-field approach always breaks down at
sufficiently long times. Also, due to the compressible transport,
strong inhomogeneities appear in the particle spatial distribution,
changing the local effective reaction rate. To describe such a system
we introduced the joint $n$-point density that was shown to obey a
hierarchy of transport equations.  This hierarchy was then solved
using a Lagrangian approach based on averaging over particle
trajectories. The solutions are shown to decay exponentially at long
times unlike the algebraic $\propto t^{- 1}$ standard mean-field
prediction. Then, it was explicitly shown that all moments of the
number of particles decrease with the same rate $\gamma$. This
effective rate depend on the statistics of the relative distance
between a pair of Lagrangian trajectories. One of the main results of
this work was to provide a formulation of the problem in terms of
non-interacting Lagrangian trajectories of the flow. This approach
allowed us to apply some standard tools of statistical mechanics and
passive (non-reacting) compressible transport to the problem of
advection-reaction. Assuming the existence of a large deviation
principle for the time that two non-reacting particles spend within a
distance $a$, we expressed the effective rate in terms of the
(Cram\'er) rate function. The asymptotic values of the effective rate
$\gamma$ for $\mu\to 0$ and $\mu\to\infty$ were obtained using general
properties of the rate function. We then introduced a phenomenological
model that allowed us to understand how the effective rate depends on
the compressibility of the carrier flow and on the interaction
radius. Finally, in the case of reactants advected by a smooth
compressible Gaussian delta-correlated-in-time flow (Kraichnan model),
an analytic solution obtained by using Feynman-Kac formula can be
written.

Many aspects of this work can be easily extended to more complicated
situations. For instance we have focused only on tracers, that is
particles that just follow a prescribed flow and whose $d$-dimensional
position space dynamics is given by equation~\eqref{eq:tracers}. This
choice was made in order to simplify the presentation. Most of the
results presented in this work remain valid when considering particles
obeying a Newton equation, that is whose dynamics is in the
$2d$-dimensional position-velocity phase space.  For instance, the
expression \eqref{Eq:AsymPnIntegrated} for the mean number of particles
in terms of relative distances remains formally the same, but one has
to adopt a gas kinetics approach and perform an additional integration
over particle velocities. Also, we considered particles reacting
within a sharp interaction radius. This hypothesis can be easily
extended replacing the Heaviside function by a more general kernel,
which could possibly depend on velocity differences or any other
intrinsic property of the particles.

Despite these straightforward extensions, many questions are still
open. For instance, we showed that factorial moments of the number of
particles exponentially decrease with a rate that is an increasing
function of their order. However, only in the zero-dimensional case an
explicit dependence was given (namely $\gamma_n=\mu
n(n-1)/2\gamma_2$). In $d\ge1$, these effective rates depend on the
joint statistics of several particle pairs. Multi-time correlations of
non-independent pairs of Lagrangian trajectories directly intervene
making difficult to evaluate the effective rates. Developing such
theoretical tools is important not only for the reaction-advection
problem but also for general questions associated with particle
transport. Another issue that was not addressed in this work,
corresponds to the case when the velocity flow is not smooth enough to
guarantee the existence and uniqueness of the Lagrangian
trajectories. This is for example, the case of turbulent flows when
the interaction radius is greater than the Kolmogorov length scale. In
such cases, even if the solution in terms of tracers could be in
principle extended by using the statistical description of the
so-called generalized Lagrangian flow \cite{gawedzki2000phase}, many
of the theoretical considerations can drastically change. For
instance, in these situations the relative dispersion of particles is
very different from the Brownian-motion-with-drift lognormal
description used in this work.

Another topic, not covered by this work, concerns events related to
the right-hand side of the rate function, which corresponds to the
rare situations where particles stay close for times much larger than
the average. This is relevant for instance when considering the
probability that particles react very fast, a situation that can lead
to underestimating the amount of reactants needed in some practical
applications.

Finally, a natural extension of this work is to consider more general
reactions. In particular, the case of coalescence/coagulation in
dilute media presents many relevant applications to different fields,
such as astrophysics or geosciences. Note that an alternative approach
to that given by the Lagrangian trajectories consists in trying to
close the hierarchy of transport equations by performing an
appropriate coarse-graining. However, because of the finiteness of the
interaction radius this approach is not straightforward.

\begin{acknowledgements}
  We are extremely grateful to one of the Referees for pointing to us
  the fact that the effective rate in Kraichnan flow could be obtained
  solving an eigenvalue problem. We acknowledge Colm Connaughton,
  Paolo Muratore-Ginanneschi, Marija Vucelja, Samriddhi S. Ray for
  many useful discussions and remarks. The research leading to these
  results has received funding from the European Research Council
  under the European Community's Seventh Framework Program
  (FP7/2007-2013, Grant Agreement no. 240579).  MC acknowledges
  Observatoire de la C\^ote d'Azur for hospitality, and the support of
  MIUR PRIN-2009PYYZM5.

\end{acknowledgements}

\bibliographystyle{spmpsci}       

\begin{thebibliography}{10}

\bibitem{Bec2004}
Bec, J., Gaw{\c e}dzki, K., Horvai, P.: Multifractal clustering in compressible
  flows.
\newblock Phys. Rev. Lett. \textbf{92}, {224{}501} (2004)

\bibitem{CardyImaginaryNoise}
Cardy, J.: Renormalisation group approach to reaction-diffusion problems.
\newblock In: J.M. Drouffe, J.B. Zuber (eds.) {The Mathematical Beauty of
  Physics: A Memorial Volume for Claude Itzykson, Saclay, France 5-7 June
  1996}. World Scientific, Singapore ; River Edge, NJ (1997)

\bibitem{chertkov2003boundary}
Chertkov, M., Lebedev, V.: Boundary effects on chaotic advection-diffusion
  chemical reactions.
\newblock Phys. Rev. Lett. \textbf{90}, 134{}501 (2003)

\bibitem{Crisanti93}
Crisanti, A., Paladin, G., Vulpiani, A.: Products of random matrices in
  statistical physics.
\newblock Springer Verlag, Berlin; New York (1993)

\bibitem{Deloubriere2002}
Deloubri\`{e}re, O., Frachebourg, L., Hilhorst, H., Kitahara, K.: {Imaginary
  noise and parity conservation in the reaction $A+A \leftrightharpoons 0$}.
\newblock Physica A \textbf{308}, 135 (2002)

\bibitem{devenish2012droplet}
Devenish, B., Bartello, P., Brenguier, J., Collins, L., Grabowski, W.,
  IJzermans, R., Malinowski, S., Reeks, M., Vassilicos, J., Wang, L., et~al.:
  Droplet growth in warm turbulent clouds.
\newblock Q. J. R. Meteorol. Soc.  (2012)

\bibitem{doi1976stochastic}
Doi, M.: Stochastic theory of diffusion-controlled reaction.
\newblock J. Phys. A \textbf{9}, 1479 (1976)

\bibitem{donsker1975asymptotic}
Donsker, M.D., Varadhan, S.S.: Asymptotic evaluation of certain markov process
  expectations for large time, i.
\newblock Communications on Pure and Applied Mathematics \textbf{28}(1), 1--47
  (1975)

\bibitem{Ellis85}
Ellis, R.: Entropy, Large Deviations, and Statistical Mechanics.
\newblock Springer-Verlag, New-York (1985)

\bibitem{falkovich2002acceleration}
Falkovich, G., Fouxon, A., Stepanov, M.: Acceleration of rain initiation by
  cloud turbulence.
\newblock Nature \textbf{419}, 151 (2002)

\bibitem{FalkovichG.2001}
Falkovich, G., Gaw{\c e}dzki, K., Vergassola, M.: {Particles and fields in
  fluid turbulence}.
\newblock Rev. Mod. Phys. \textbf{73}, 913 (2001)

\bibitem{FMNV99}
{Frisch}, U., {Mazzino}, A., {Noullez}, A., {Vergassola}, M.: {Lagrangian
  method for multiple correlations in passive scalar advection}.
\newblock Phys. Fluids \textbf{11}, 2178 (1999)

\bibitem{gardiner1977poisson}
Gardiner, C., Chaturvedi, S.: The poisson representation. i. a new technique
  for chemical master equations.
\newblock J. Stat. Phys. \textbf{17}, 429 (1977)

\bibitem{Gardiner_1985}
Gardiner, C.W.: Handbook of Stochastic Methods for Physics, Chemistry and the
  Natural Sciences.
\newblock Springer-Verlag, Berlin; New York (1985)

\bibitem{gawkedzki2004sticky}
Gaw{\c e}dzki, K., Horvai, P.: Sticky behavior of fluid particles in the
  compressible kraichnan model.
\newblock Journal of statistical physics \textbf{116}(5-6), 1247--1300 (2004)

\bibitem{gawedzki2000phase}
Gaw{\c e}dzki, K., Vergassola, M.: Phase transition in the passive scalar
  advection.
\newblock Physica D \textbf{138}, 63 (2000)

\bibitem{goudon2004homogenization}
Goudon, T., Poupaud, F.: Homogenization of transport equations: weak mean field
  approximation.
\newblock SIAM J. on Mathematical Analysis \textbf{36}, 856 (2004)

\bibitem{grassberger1988}
Grassberger, P., Badii, R., Politi, A.: Scaling laws for invariant measures on
  hyperbolic and nonhyperbolic atractors.
\newblock J. Stat. Phys. \textbf{51}, 135 (1988)

\bibitem{johansen2007rapid}
Johansen, A., Oishi, J., Mac~Low, M., Klahr, H., Henning, T., Youdin, A.: Rapid
  planetesimal formation in turbulent circumstellar disks.
\newblock Nature \textbf{448}, 1022 (2007)

\bibitem{KacLemma}
Kac, M.: On the notion of recurrence in discrete stochastic processes.
\newblock Bull. Amer. Math. Soc. \textbf{53}, 10{}024 (1947)

\bibitem{mattis1998uses}
Mattis, D., Glasser, M.: The uses of quantum field theory in diffusion-limited
  reactions.
\newblock Rev. Mod. Phys. \textbf{70}, 979 (1998)

\bibitem{Maxey2006}
Maxey, M.R.: {The gravitational settling of aerosol particles in homogeneous
  turbulence and random flow fields}.
\newblock J. Fluid Mech. \textbf{174}, 441 (1987)

\bibitem{oksendal1998stochastic}
{\O}ksendal, B.: Stochastic differential equations.
\newblock Springer (1998)

\bibitem{paladin1987anomalous}
Paladin, G., Vulpiani, A.: Anomalous scaling laws in multifractal objects.
\newblock Phys. Rep. \textbf{156}, 147 (1987)

\bibitem{peliti1985path}
Peliti, L.: Path integral approach to birth-death processes on a lattice.
\newblock J. Physique \textbf{46}, 1469 (1985)

\bibitem{Perlekar2010}
Perlekar, P., Benzi, R., Nelson, D., Toschi, F.: {Population Dynamics At High
  Reynolds Number}.
\newblock Phys. Rev. Lett. \textbf{105}, 144{}501 (2010)

\bibitem{pruppacher1998microphysics}
Pruppacher, H., Klett, J.: Microphysics of clouds and precipitation.
\newblock Kluwer Academic Publisher, Dordrecht (2010)

\bibitem{shaw2003particle}
Shaw, R.: Particle-turbulence interactions in atmospheric clouds.
\newblock Ann. Rev. Fluid Mech. \textbf{35}, 183 (2003)

\bibitem{Tauber2005}
T\"{a}uber, U.C., Howard, M., Vollmayr-Lee, B.P.: {Applications of
  field-theoretic renormalization group methods to reaction--diffusion
  problems}.
\newblock J. Phys. A \textbf{38}, R79 (2005)

\bibitem{Neufeld2007}
Torney, C., Neufeld, Z.: Transport and aggregation of self-propelled particles
  in fluid flows.
\newblock Phys. Rev. Lett. \textbf{99}, 78{}101 (2007)

\bibitem{vanKampen1992}
Van~Kampen, N.: Stochastic processes in physics and chemistry.
\newblock Elsevier, Amsterdam (1992)

\bibitem{vergassola1997scalar}
Vergassola, M., Avellaneda, M.: Scalar transport in compressible flow.
\newblock Physica D \textbf{106}, 148 (1997)

\end{thebibliography}

\end{document}